\documentclass[conference]{IEEEtran}
\IEEEoverridecommandlockouts
\usepackage{cite}
\usepackage{amsmath,amssymb,amsfonts}
\usepackage{algorithmic}
\usepackage{graphicx}
\usepackage{textcomp}
\usepackage{xcolor}
\def\BibTeX{{\rm B\kern-.05em{\sc i\kern-.025em b}\kern-.08em
T\kern-.1667em\lower.7ex\hbox{E}\kern-.125emX}}
\usepackage{tikz}
\usepackage{amsmath}
\usepackage{amsmath,amsfonts, amsthm}
\usepackage{algorithmic}
\usepackage{graphicx}
\usepackage{textcomp}
\usepackage{xcolor}
\usepackage{mathtools}
\usepackage{booktabs}
\usepackage{tabularx}
\usepackage{multirow}
\usepackage{msc}
\usepackage{pgfplots}
\usepackage{etoolbox}
\usepackage{soul}

\newtheorem{definition}{Definition}[section]

\newtoggle{review}
\togglefalse{review}
\iftoggle{review}{%
    \newcommand{\x}[2]   {\textcolor{#1}{#2}}   %
    \newcommand{\stx}[1] {\st{{#1}}}              %
}{%
    \newcommand{\x}[2]   {}                     %
    \newcommand{\stx}[1] {}                     %
}

\begin{document}

\bstctlcite{MyBSTcontrol}

\title{Secure Traversable Event logging for Responsible Identification of Vertically Partitioned Health Data
}

\author{\IEEEauthorblockN{Sunanda Bose}
\IEEEauthorblockA{
\textit{Simula Research Laboratory}\\
Oslo, Norway \\
sunanda@simula.no}
\and
\IEEEauthorblockN{Dusica Marijan}
\IEEEauthorblockA{
\textit{Simula Research Laboratory}\\
Oslo, Norway \\
dusica@simula.no}
}

\maketitle

\begin{abstract}
We aim to provide a solution for the secure identification of sensitive medical information. We consider a repository of de-identified medical data that is stored in the custody of a Healthcare Institution. The identifying information which is stored separately can be associated with the medical information only by a subset of users referred to as custodians. This paper intends to secure the process of associating identifying information with sensitive medical information. We also enforce the responsibility of the custodians by maintaining an immutable ledger documenting the events of such information identification. The paper proposes a scheme for constructing ledger entries that allow the custodians and patients to browse through the entries which they are associated with. However, in order to respect their privacy, such traversal requires appropriate credentials to ensure that a user cannot gain any information regarding the other users involved in the system unless they are both involved in the same operation.
\end{abstract}

\begin{IEEEkeywords}
Security and Privacy, Healthcare, Accountable private information retrieval
\end{IEEEkeywords}

\section{Introduction}
\label{sec:introduction}
Although Health Data (HD) is generally considered private information owned by patients, it is often stored in the custody of a healthcare institute.
HD may be used for medical research involving researchers inside or outside the jurisdiction of the institute.
The institutes generally anonymize the data before sending them to researchers working outside the jurisdiction.
However, the institute can also have some researchers working under the jurisdiction of the institute.
The internal researchers may use the data to find out the incompleteness and inconsistencies in the medical record \cite{Chatterjee2016}.
Incompleteness may signify that some documents (e.g. laboratory tests or doctor's reports) that are supposed to be in the repository are not yet submitted.
If any inconsistency is spotted by the internal researchers, this may lead to the decision of repeating some clinical tests or can even identify a misdiagnosis.

National healthcare systems of several countries also maintain documentation of the population\cite{Laugesen2021}\cite{Chatterjee2016}\cite{Bouchardy2014}. These registries comprise different types of health data, associated with their personal identity numbers \cite{Laugesen2021}. The Cancer Registries usually collect data from various sources, like hospitals, clinicians, dentists, laboratories, radiotherapy data, Death Certificates\cite{Pukkala2018}. In \cite{Chatterjee2016} and \cite{Chaudhry2002}, two methods of data collection are suggested. In the passive method, the institutions send information to the Cancer registries.
In the active method, the staff from the Cancer Registry collects the data from these institutions. A team of Registry personnel working in the jurisdiction ensures the quality of the data by checking duplicate entries and validating the consistency of records. Such teams are typically led by a medically-qualified Principal Investigator who has a background in epidemiology and/or public health~\cite{Chatterjee2016}. These registries maintain a group of internal experts, who regularly analyze HD stored in their registry \cite{Chatterjee2016}.
There can be multiple medical records associated with a patient.
This can be considered as a one-to-many relationship between identifying information and medical information.
As these data contain highly sensitive information, it has to be protected against adversarial access.
Only legitimate users of these data are the internal experts, referred to as \emph{Custodians}, who may identify the patient associated with a medical record.

Hence, we summarize legitimate data access scenarios.

\begin{enumerate}
	\item Identify a patient associated with a record.
	\item Fetch or insert data associated with  a patient.
\end{enumerate}

The custodian performing these operations is gaining significant private information about the patient.
Although the legal framework permits the custodian to gain that information, it has to be ensured that the gained information is not used for malicious purposes.
However, the events of the utilization of that information for malicious purposes may only happen after the event of information gain has happened.
In the case of malicious usage, the events of information gain can be correlated only if those events are logged.
Such an approach can promote the legitimacy of information gain by ensuring the responsibility of the custodians.
Although there have been research works addressing the security and privacy concerns of private information storage and retrieval, which is mentioned in Section~\ref{sec:related}, these works do not address the problem of responsible identification of de-identified sensitive data.

Therefore, this paper proposes a secure system of storage and retrieval of HD that can be accessed by custodians with sufficient credentials.
As such access can lead to information gain about the patients, the events of such access are documented in an immutable ledger which can be securely traversed by the custodians and the patients with appropriate credentials.
However, in order to design such a solution we have to overcome some technical challenges.
The ledger has to be protected from adversarial access to ensure the privacy of the custodians as well as the patients.
Simultaneously, the legitimate users, (custodians and the patients) should be permitted to traverse through the ledger and analyze the events of information gain that relates to them.
Moreover, the supervisor(s) (often termed as Principal Investigator\cite{Chatterjee2016}) may need to access the ledger to correlate the events with some malicious indecent and verify its legitimacy.
We also evaluate our proposed solution in terms of security and performance.

The paper is organized as follows.
A brief summary of existing works related to our problem is presented in Section~\ref{sec:related}. 
In section~\ref{sec:problem} we formulate the scientific problem of responsible identification and present the functional requirements.
We present our proposed solution in Section~\ref{sec:solution}.
The security and performance of our proposed solution is evaluated in Section~\ref{sec:evaluation}.
Finally, Section~\ref{sec:conclusion} concludes the paper.

\section{Related Work}
\label{sec:related}
Ensuring the privacy of patients’ sensitive data is an essential requirement of managing HD \cite{bose2023}. To ensure the confidentiality of HD, authors in \cite{XIA2017} implemented an AT\&T-based scheme for access control of medical records. In \cite{Gajanayake2014}, the authors describe several access control mechanisms (RBAC, MAC, DAC, and PBAC) and their applicability for ensuring the privacy of the HD. However, restricting access to the documents is not our only objective. We want to make the user responsible for accessing the document. Moreover, encrypted documents are difficult to search for or analyze. We only need de-identified data that can be used for knowledge discovery without directly revealing the patient's identity.

Threshold cryptosystem has been used in literature \cite{Thummavet2013}\cite{Jose2013}\cite{Eskeland2007} to provide shared access to health data. Cryptographic access schemes like IBE, CP-ABE are used in \cite{Yuliana2017}, \cite{Sudarsono2017} \cite{Thummavet2013} \cite{Liu2018}. In \cite{Sudarsono2017} medical data is first encrypted by the sender using the symmetric encryption algorithm AES. The secret key is encrypted using IBE and shares the encrypted document along with the encrypted key. In \cite{Yuliana2017} a Hierarchical access scheme is proposed, where the Public Health Office serves as the Public Key Generator (PKG) at the highest level, and the Hospitals, and Clinics are at the lower level. The storage servers located at hospitals and clinics store the medical records of their patients only. The public storage server is responsible for storing the referral medical records. In \cite{Ge2020} IBE is used along with a Markle Hash Tree to ensure the deletion of HD.

Our problem also requires the HD to be shared among multiple custodians. 
However, our intention is to allow one custodian to access the patients' records independently without any co-operations from other custodians.

Vertical Partitioning of data is a popular technique of de-identifying data which is often used along with anonymization.
In \cite{Yang2015} HD is partitioned into three tables.
One contains the original medical information, without the identifying attributes. The other two tables contain anonymized quasi-identifiable\footnote{The attributes that can reveal important information about the identity of the patient when correlated with publicly available information.} attributes and encrypted ciphertexts of identifying and quasi-identifiable attributes. Different healthcare institutions may maintain records as vertically partitioned data\cite{Domadiya2021} which may be mined for scientific or statistical purposes. Data anonymization techniques are often used to protect medical data from being re-identified when correlated with publicly available information. However, it processes the original data and generalizes the values of attributes which reduces the amount of information\cite{Oh2021}. Such techniques include k-anonymity\cite{Li2018}\cite{Yang2015}, l-diversity\cite{Machanavajjhala2006}, t-closeness\cite{Li2011t} etc.. are often used by healthcare registries while exchanging HD with external research institutions. However, our objective is to allow the identification of HD while ensuring accountability.

Blockchain-based techniques are often used for HD-related transactions ~\cite{9963549}, ~\cite{Huang2020}, ~\cite{Tian2019}, although they bring significant challenges related to validating the correctness of such solutions \cite{M2022100492}. In \cite{Tian2019} all transactions are performed using smart contracts that provide two functions, store and get, and all data is stored in the blockchain as key-value pairs. In \cite{Huang2020} the patients may delegate hospitals to encrypt their medical records and store them on semi-trusted cloud servers. However, in our case, the patients are not actively participating in the process. Rather, the patients remain passive while the events of their records being accessed get documented which can be viewed by them later.

\begin{table}[]
	\centering
	\begin{tabularx}{\linewidth}{|l|X|}
		\toprule
		Symbol & Usage                                \\ \midrule
		$p, q, g$					   & Modulus, Subgroup order and generator of the group. \\
		$\pi_{x}, y_{x} = g^{\pi_{x}}$ & Private and Public key of user $A_{x}$ \\
		$Y = \bigcup\limits_{x\in X} y_{x}$  & Set of public keys of all users $X$. \\ \midrule
		$\xi,\zeta$                      & Symmetric Encryption, Decryption Algorithm. 
  \\
		$H,H_{2}$                        & Cryptographic Hash functions e.g. SHA512 \\
		$x \in_{\mathbb{R}} X$     & $x$ is a random integer from set $X$ \\ \midrule
		$\frac{a}{b}$              & $a b^{\prime}$ where $b^{\prime}$ is the multiplicative inverse of $b$ in $\mathcal{Z}_{p}$, such that $b b^{\prime} \equiv 1 \; (mod\, p)$ \\
		$a^{-1}$                   & Multiplicative inverse of $a$ in $\mathcal{Z}_{(p-1)}$, such that $a a^{-1} \equiv 1 \; (mod\, (p-1))$ \\ \midrule
		$f(x) \rightarrow y$       & $y$ is deterministically computable using $f$ and $x$. \\ \midrule
		$\tau_{x}^{(k)}$           & $k^{th}$ block in which user $A_{x}$ was active \\
		$\tau_{\widehat{x}}^{(k)}$ & $k^{th}$ block in which user $A_{x}$ was passive \\
		$\tau^{(r)} $              & $r^{th}$ block in the ledger, where the information regarding the involvement of any user is either irrelevant or unknown. \\
		\bottomrule
	\end{tabularx}
	\caption{Symbol Table}
	\label{table:symbol}
\end{table}

\section{Responsible Record Identification Problem}
\label{sec:problem}
For privacy-related concerns, the database is often vertically partitioned where personal information is separated from the sensitive medical information\cite{Yang2015}.
In our proposed system, we assume a medical record is de-identified and the sensitive information is stored separately from the identifying information.
Only the permitted users can identify the patient associated with that de-identified record and can also find all medical records associated with the patient.
We may refer to this action as \emph{record identification}.
However, there are two constraints applied to the identification operations.
First, even if some adversary gets access to the storage server, it should not be possible to perform any of these identification actions without the participation of the permitted users.
Second, in order to ensure the responsibility of the identifier, an entry must be created in the immutable ledger whenever each of these actions is performed.

We refer to these actions of record identification as an \textit{Access Event}.
An Access Event is participated by two users.
The permitted user that initiates the identification process is considered \textit{active} because this user is actively communicating in order to identify the record for further analysis.
The patient whose record is being accessed may be unaware of such an event before it happens.
The decision that the records associated with that patient have to be accessed may not be taken in the active participation of that patient.
Hence we consider the patient as a \textit{passive} user.
We need a ledger that documents the access events between these two users.
Both of these users should be allowed to browse through the events logged into the ledger securely.
The event participation information must not be disclosed to anyone other than these active and passive users and the supervisors who have sufficient credentials to access any random entry and view participation information.
The active user is allowed to navigate to the next and the previous entries in which the same user was active.
Similarly, the passive user is allowed to navigate to the next and the previous entries in which the same user was passive.
But no users should be allowed to navigate to ledger entries associated with a different user.
We formulate these problems in Section~\ref{sec:problem-storage} and \ref{sec:problem-block}
(the symbols used in the formulation and throughout the paper are shown in Table~\ref{table:symbol}).
Additionally, in order to ensure the immutability of our ledger, each entry contains the cryptographic hash of its previous entry, in the order of time, which may not be participated by the same users.

\subsection{Secure Storage}
\label{sec:problem-storage}

\begin{figure}
	\centering
	\includegraphics[width=0.76\linewidth]{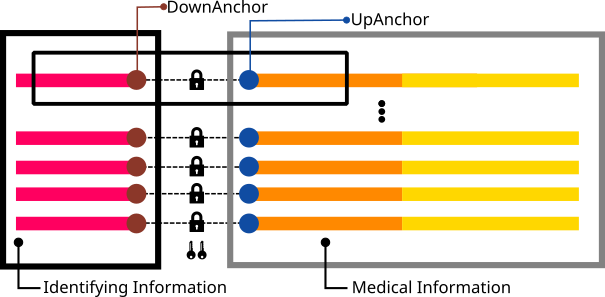}
	\caption{Secure De-Identification}
	\label{fig:secure-de-identification}
\end{figure}

To secure record identification, we associate each side of the partitioned record with a ciphertext that we refer to as an \textit{anchor}.
In  Figure~\ref{fig:secure-de-identification} the identifying anchor is shown in brown while the sensitive information anchor is shown in blue.
Formally, given a complete record $D_{i}$, the set of personal information $dp_{i}$ and the set of medical information $dm_{i}$ are annotated with different anchors $dp_{i}^{*}, dm_{i}^{*}$ respectively. However, $\exists\; \overrightarrow{f}(dp_{i}^{*}, x) \rightarrow dm_{i}^{*}, \overleftarrow{f}(dm_{i}^{*}, x) \rightarrow dp_{i}^{*}$ can relate both of these anchors, where $x$ is the secret that can be obtained securely by the cooperation of permitted users only.
For fast retrieval, the records are indexed with $dp_{i}^{*}, dm_{i}^{*}$ .
\subsection{Secure Traversable Ledger}
\label{sec:problem-block}

Given two ordered entries in the ledger $\tau_{u}^{(k)}, \tau_{u}^{(k+1)}$ in which user $A_{u}$ is active $\exists \; \overrightarrow{f_{a}}, \overleftarrow{f_{a}} $ such that  $\overrightarrow{f_{a}}(\tau_{u}^{(k)}, \pi_{u}) \rightarrow \tau_{u}^{(k+1)}$ and $\overleftarrow{f_{a}}(\tau_{u}^{(k+1)}, \pi_{u}) \rightarrow \tau_{u}^{(k)}$ where $\pi_{u}$ is a secret that only $A_{u}$ has access to.
Similarly, if user $A_{v}$ is passive and $\tau_{v}^{(k)}, \tau_{v}^{(k+1)}$ are two ordered entries in which it was passive, then $\exists \; \overrightarrow{f_{p}}, \overleftarrow{f_{p}}$ such that  $\overrightarrow{f_{p}}(\tau_{v}^{(k)}, \pi_{v}) \rightarrow \tau_{v}^{(k+1)}$ and $\overleftarrow{f_{p}}(\tau_{v}^{(k+1)}, \pi_{v}) \rightarrow \tau_{v}^{(k)}$.
In this paper $\overrightarrow{f_{a}},  \overleftarrow{f_{a}}$ are referred to as active forward and backward functions while $\overrightarrow{f_{p}}, \overleftarrow{f_{p}}$ as passive forward and backward traversal functions.
We also refer to these four functions as \emph{traversal requirements} that our proposed solution has to satisfy.

In order to make the traversal secure $\nexists \overrightarrow{f^{\prime}_{a}}(\tau_{u}^{(k)}, x) \rightarrow \tau_{u}^{(k+1)}$ such that $x \neq \pi_{u}$, and same applies for the other traversal functions. Also there $\nexists\; F_{a}(\tau_{u}^{(k)}, \tau_{u}^{(k+1)}) \rightarrow [0,1], F_{p}(\tau_{v}^{(k)}, \tau_{v}^{(k+1)}) \rightarrow [0,1]$ that deterministically produces a binary output denoting the given two entries are related to the same active or passive user respectively.
In that case, an adversary can apply that function on all pairs or entries to partition all entries belonging to the same user. 

\begin{figure}
	\centering
	\includegraphics[width=0.58\linewidth]{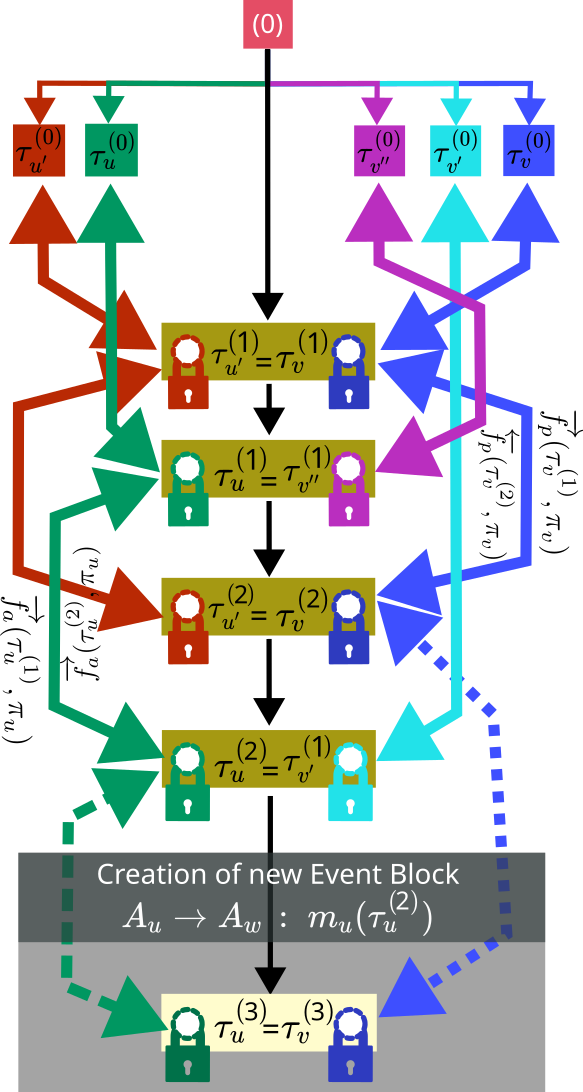}
	\caption{Traversable Ledger}
	\label{fig:ledger}
\end{figure}

In this paper, we refer to these entries as \emph{Block}.
In Figure~\ref{fig:ledger}, we show 4 blocks (shown in yellow rectangles) each referring to an Access Event.
In the first block from the top, user $A_{u^{\prime}}$ is active while the user $A_{v}$ is passive and it is the first Access Event associated with both of these users.
Similarly, in the fourth block, the user $A_{u}$ is active while the user $A_{v^{\prime}}$ is passive.
The user $A_{u}$ can reach this block using the function $\overrightarrow{f_{a}}$ and its private credentials, which is reachable from $\tau_{u}^{(0)}$.
Although it is the fourth block in the order of time, it is the second block that $A_{u}$ can jump into through active traversal.
Similarly, the user $A_{v^{\prime}}$ can jump into this block by traversing only once from $\tau_{v^{\prime}}^{(0)}$.
We label these blocks from these users' perspectives.
Hence, the same block is referred to as $\tau_{u}^{(2)}$ and $\tau_{v^{\prime}}^{(1)}$ by users $A_{u}$ and $A_{v^{\prime}}$.
Genesis blocks are shown on the top, which are the first blocks ($0^{th}$) associated with each user.

Each user is associated with a dedicated chain of events, that has a traversable total order.
However, a block in a user-specific chain may overlap with some other user's chain.
As these chains are sets of blocks, the ledger proposed in our work can be described as a union of totally ordered sets.
All these totally ordered sets start with a genesis block that does not have a mutual order in the context of the user's secret, which makes this union a partially ordered set.
However, all blocks, including the genesis blocks, are totally ordered with respect to time.
Although the order of these sets is secret in the absence of users' secret, the total order of the ledger, which is the union of all these sets, is transparent as it is ordered by time.
Additionally, ensuring the total order of the blocks inside the ledger implies that the order of the sets is also maintained.
The colored arrows in the figure denote active or passive traversals.
In the end, the creation of a new block is shown in the figure, which can be traversed from the last blocks $\tau_{u}^{(2)}$ and $\tau_{v}^{(2)}$.

\section{Proposed Solution}
\label{sec:solution}

The human entities in the problem are the Data Managers and Supervisors and patients.
In this paper, we often refer to the Data Managers and Supervisors as Custodians because they are in the jurisdiction of the Institution.
A Custodian is permitted to access and identify the medical record(s) without depending on any other custodian.
The responsibility is enforced by creation of an entry in the immutable ledger.
The Database is vertically partitioned but not encrypted.
Hence any unauthorized user, if got access to the Database may see the medical records but will not have the ability to identify the patient associated with the medical record.
We choose to keep the data unencrypted to permit fast access and query processing which is often required for medical research.
The focus of this work is to make the Custodians responsible for record identification.
So, we do not consider the data thefts as long as the stolen data still remains non-identifiable.
With this setup, an adversary with database access may still use statistical techniques to re-identify the vertically partitioned data.
However, our paper does not deal with re-identification threats.
The focus of our paper is concentrated on enforcing the accountability of identification.

\subsection{System Design}
\label{sec:system-design}
The proposed solution consists of Custodians, Patients, a Trusted Server, a Database, and a Key-value store that serves the purpose of the ledger. 
Only the Trusted Server (TS) has access to the database stored on a different Server.
The events are logged in the key-value store accessible by the TS and all other human entities.
A Custodian performs operations like \emph{Insertion}, \emph{Identification}, etc. on the de-identified HD through the TS.
After each operation, an entry is written into the Ledger.
The ledger consists of entries referred to as blocks, each associated with an Access Event.
Each block contains the cryptographic hash of its previous block in the ledger.
With appropriate credentials, the users can selectively browse through the blocks that are associated with them.
The TS is the only entity that writes into the ledger.
Anyone including the custodians and the patients can read from the ledger. 
We follow a semi-honest adversarial model for the TS, which implies that the TS follows the protocol when interacting with the custodians.
However, the TS can be curious,  to explore the database with an intention to identify some medical records while not interacting with any custodians.
Our scheme requires the cooperation of two entities in order to identify a record, one of which is the TS and the other is a custodian.
Both custodians and patients can traverse the ledger and read the blocks related to them.
We consider the custodians and patients to be malicious.
They may deviate from the protocol to gain information about other custodians or patients.

We use Diffie Hellman\cite{Diffie1976} based construction for the anchors and the \emph{blocks}. 
Hence, security is based on the assumption that the adversary cannot solve CDH and DDH problems defined in Definition \ref{def:cdh} and \ref{def:ddh}.
Each actor (Custodians and TS) $A_{t}$ in our system has a pair of private key $t$ and public key $g^{t}$ generated by the TS, such that $\exists t^{-1}: g^{tt^{-1}} \equiv g\, (mod p)$.
We assume that the key generation and distribution process is secure.
In this paper the symbols $A_{u}, A_{v}, A_{w}$ are used for denoting an Active user (Data Managers and Supervisors), Passive user (Patients) and the TS, respectively.  
We use the symbol $A_{s}$ to specifically denote a supervisor.

\begin{definition}
	Given a cyclic group $G$ of order $q$, with generator $g$, and $\{g^{a},g^{b}\}$ Computational Diffie Hellman (CDH) problem is to compute $g^{ab}$ where $a,b \in_{R} Z_{q}^{*}$.
	\label{def:cdh}
\end{definition}
\begin{definition}
	Given a cyclic group $G$ of order $q$, with generator $g$, and $\{g^{a},g^{b}\}$ Decisional Diffie Hellman (DDH) problem is to distinguish $g^{ab}$ from $g^{c}$ where $a,b,c \in_{R} Z_{q}^{*}$.
	\label{def:ddh}
\end{definition}

\subsection{Securely Identifiable Vertical Partitioning}
\label{sec:db-partitioning}

\begin{figure}
        \centering
	\includegraphics[width=0.8\linewidth]{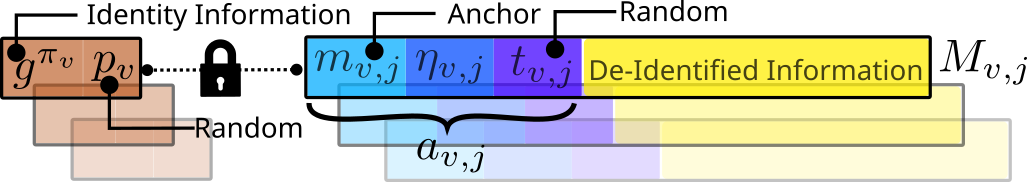}
	\caption{Database Anchors}
	\label{fig:db-anchor}
\end{figure}

In Figure~\ref{fig:db-anchor}, the formulation of a vertically partitioned record is shown.

The medical information parts of these records $\{M_{v,1}, M_{v,2}, \dots\}$ do not include any identifying information.
The Identifying information (on the left) and the De-Identified information (on the right) are stored in different tables.
The identifying information part contain a random number $p_{v} \in_{R} [1, 2^{512}]$ indexed by the public key $g^{\pi_{v}}$ of the patient.
Each medical record $M_{v,j}$ is associated with an anchor $a_{v,j}$ expressed as a tuple $\{m_{v,j}, \eta_{v,j}, t_{v,j}\}$ indexed by $m_{v,j}$.
The construction of anchor $a_{v,j}$ is shown in Equation~\ref{eq:identify-anchor}.

The system is initialized with $\theta \in_{\mathbb{R}} \mathbb{Z}_{p}$ such that $\nexists \; \theta^{-1} \in \mathbb{Z}_{p}$, but for its hash $h = H(g^{\theta})$, $\exists h^{-1}$ such that $g^{hh^{-1}}\; \equiv g\, (mod\; p)$.

\begin{equation}
	\begin{split}
		m_{v,j} &= \xi\Big(g^{\pi_{v}}, H_{2} (g^{\theta t_{v,j-1}})\Big) \\
		\eta_{v,j} &= t_{v,j-1} H(g^{\theta t_{v,j}})
	\end{split}
	\label{eq:identify-anchor}
\end{equation}

With knowledge of $t_{v,j}$ and $g^{\theta}$, one can compute $H_{2}(g^{\theta t_{v,j}})$ which is used as the key in the symmetric encryption algorithm $\xi$.
Hence, with that knowledge, it is also possible to decrypt $m_{v,j+1}$ and extract $g^{\pi_{v}}$ and identify the patient.
$t_{v,j}$ a random number stored as a plaintext in $a_{v,j}$.
Similarly, one can also extract $t_{v,j}$ from $\eta_{v,j+1}$ by computing the hash of $(g^{\theta})^{t_{v,j+1}}$.
However, to iterate through all medical records of a patient, we need a first record to start with.
However, the first record needs a $t_{v, -1}$ which does not exist.
Hence, we use $p_{v}$ associated with the identifying part of the record. 
Therefore, we can relate any patient and sensitive information of that patient as long as $g^{\theta}$ is known.

However, $\theta$ or $g^{\theta}$ is not remembered or stored in any persistent storage. 
Rather, an access key is computed using that and the user's private key 

as $(g^{\pi_{t}})^{\theta w} = g^{\pi_{t}\theta w}$ %

A user $A_{u}$ can use the multiplicative inverse of its private key $\pi_{u}$ and send that to the Trusted server who can use the multiplicative inverse of $w$ to recover the $g^{\theta}$ which was previously lost in the beginning as shown in Equations~\ref{eq:access-key-recovery}.

\begin{equation}
	\begin{split}
		(g^{\pi_{u}\theta w})^{\pi_{u}^{-1}} = g^{\theta w} \Rightarrow (g^{\theta w})^{w^{-1}} = g^{\theta}
	\end{split}
	\label{eq:access-key-recovery}
\end{equation}

However, we need to make the exchange in a way that enforces the creation of an entry in the immutable ledger.
So, the exchange of shared secrets is incorporated into a protocol.
We call that process Request for Sensitive Information (RSI).
The RSI contains the type of operation the active user intends to perform (e.g. identify, insert, fetch etc..) and the related data.
The other is when the custodians want to identify the patients associated with a set of medical records.
In the next Section, we formulate the entries in the immutable ledger.
In Section~\ref{sec:ledger-construction} we explain the construction of the entries, and then we describe the protocol that integrates ledger construction and exchange of access key.

\subsection{Ledger Formulation}
\label{sec:ledger-formulation}
\begin{figure}
    \centering
    \includegraphics[width=0.74\linewidth]{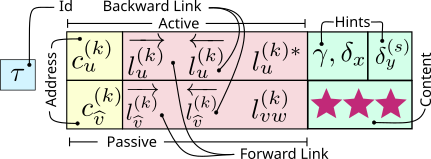}
    \caption{Block Formulation}
    \label{fig:block-formulation}
\end{figure}

d Traversability Requirements respectively.
The ledger entries, referred to as blocks, are constructed as an effect of the Access Event, which happens in collaboration with the TS which is supposed to enforce the creation of the block.
Hence, the TS constructs the block and posts it on the ledger.
However, the blocks have to be constructed in a way that satisfies the \emph{traversal requirements} described in Section~\ref{sec:problem}.
Every block has four parts, \emph{Address}, \emph{Active}, \emph{Passive}, and \emph{Content} as shown in Figure~\ref{fig:block-formulation}.
The formulations of the first three parts are presented in Equation~\ref{eq:address-formulation}, \ref{eq:active-link-formulation} and \ref{eq:passive-link-formulation}.
Each block is formulated using two random numbers $r_{u}^{(k)}, r_{v}^{(k)}$ that are cryptographically associated with the active and the passive parts of the block.
The \emph{Address} component $\{c_{u}^{(k)}, c_{\widehat{v}}^{(k)}\}$ is used to \emph{index} each block for fast retrieval.
The Active and Passive parts contain information for the Active and Passive Traversals in a forward $\{\overrightarrow{l_{u}^{(k)}}, \overrightarrow{l_{\widehat{v}}^{(k)}}\}$ and backward $\{\overleftarrow{l_{u}^{(k)}}, \overleftarrow{l_{\widehat{v}}^{(k)}}\}$ direction.
In the next section, we discuss how the \emph{traversal requirements} are satisfied by this formulation.

\resizebox{0.99\linewidth}{!}{%
\begin{subequations}
    \hspace{-7mm}
    \begin{tabularx}{1.116\linewidth}{@{}X@{}X@{}}
		\begin{equation}
			c_{u}^{(k)} = \tau_{u}^{(k-1)}H\Big(g^{\pi_{u}r_{u}^{(k-1)}}\Big) \label{eq:block-addresses-active}
		\end{equation}%
        &
		\begin{equation}
			c_{\widehat{v}}^{(k)} = \tau_{\widehat{v}}^{(k-1)}H\Big(g^{\pi_{v}r_{v}^{(k-1)}}\Big) \label{eq:block-addresses-passive}
		\end{equation}
    \end{tabularx}
    \label{eq:address-formulation}
\end{subequations}
}
\resizebox{0.97\linewidth}{!}{%
\begin{minipage}{0.52\linewidth}
    \begin{subequations}
        \begin{align}
            \begin{split}
                \overrightarrow{l_{u}^{(k)}} =& g^{r_{u}^{(k)}} \label{eq:block-active-forward}
            \end{split} \\
            \begin{split}
                \overleftarrow{l_{u}^{(k)}} =& H(g^{\pi_{u}r_{v}^{(k)}}) g^{r_{u}^{(k-1)}}  \label{eq:block-active-backward}
            \end{split} \\
            \begin{split}
                l_{u}^{(k)*} =& H\Big( g^{w\pi_{u}r_{u}^{(k)}}g^{\pi_{u}} \Big)  \label{eq:block-active-checksum}
            \end{split}
        \end{align}
        \label{eq:active-link-formulation}
    \end{subequations}
\end{minipage}%
\begin{minipage}{0.51\linewidth}
    \begin{subequations}
        \begin{align}
            \begin{split}
                \overrightarrow{l_{\widehat{v}}^{(k)}} =& g^{r_{v}^{(k)}} \label{eq:block-passive-forward}
            \end{split} \\
            \begin{split}
                \overleftarrow{l_{\widehat{v}}^{(k)}} =& H(g^{\pi_{v} r_{u}^{(k)}}) g^{r_{v}^{(k-1)}} \label{eq:block-passive-backward}
            \end{split} \\
            \begin{split}
                l_{vw}^{(k)} =& H(g^{wr_{v}^{(k)}}) g^{h\pi_{v}r_{v}^{(k)}} \label{eq:block-passive-server}
            \end{split}
        \end{align}
        \label{eq:passive-link-formulation}
    \end{subequations}
\end{minipage}
}

\subsection{Traversal}
The objective of the \emph{active forward traversal} function $\overrightarrow{f_{a}}(\tau_{u}^{(k-1)}, \pi_{u})$ is to make it possible to compute $c_{u}$ using private key of $A_{u}$, as shown in Equation~\ref{eq:traversal-active-forward}. This $c_{u}$ is then looked up in the index to find the block id $\tau_{u}^{(k)}$ and then read the block. However, for the \emph{active backward traversal}, the function $\overleftarrow{f_{a}}(\tau_{u}^{(k)}, \pi_{u})$ computes the id of the previous active block without requiring looking up in the index of the addresses.
The traversal function $\overleftarrow{f_{a}}(\tau_{u}^{(k)}, \pi_{u})$ is modeled as shown in Equation~\ref{eq:traversal-active-backward}. 

The \emph{passive forward traversal} and \emph{passive backward traversal} are very similar to the active one, however instead of using $\pi_{u}$, it requires $\pi_{v}$ which is the private key of the passive user $A_{v}$. The traversal functions $\overrightarrow{f_{p}}$ and $\overleftarrow{f_{p}}$ are shown in Equation~\ref{eq:traversal-passive-forward} and \ref{eq:traversal-passive-backward} respectively.

\begin{subequations}
	\begin{align}
		\begin{split}
			\overrightarrow{f_{a}}(\tau_{u}^{(k-1)}, \pi_{u}) &= \tau_{u}^{(k-1)} H\Bigg(\Big(\overrightarrow{l_{u}^{(k-1)}}\Big)^{\pi_{u}}\Bigg) 
			= c_{u}^{(k)} \label{eq:traversal-active-forward}
		\end{split} \\
		\begin{split}
			\overleftarrow{f_{a}}(\tau_{u}^{(k)}, \pi_{u}) &= \frac{c_{u}}{H\Bigg(\Bigg(\frac{\overleftarrow{l_{u}^{(k)}}}{H\Big(\Big(\overrightarrow{l_{v}^{(k)}}\Big)^{\pi_{u}}\Big)}\Bigg)^{\pi_{u}}\Bigg)} 
			= \tau_{u}^{(k-1)} \label{eq:traversal-active-backward}
		\end{split}
	\end{align}
\end{subequations}
\begin{subequations}
	\begin{align}
		\begin{split}
			\overrightarrow{f_{p}}(\tau_{\widehat{v}}^{(k-1)}, \pi_{v}) &= \tau_{\widehat{v}}^{(k-1)} H\Bigg(\Big( \overrightarrow{l_{\widehat{v}}^{(k-1)}} \Big)^{\pi_{v}} \Bigg) 
			= c_{\widehat{v}}^{(k)} \label{eq:traversal-passive-forward}
		\end{split} \\
		\begin{split}
			\overleftarrow{f_{p}}(\tau_{\widehat{v}}^{(k)}, \pi_{v}) &= \frac{c_{\widehat{v}}}{H\Bigg(\Bigg(\frac{\overleftarrow{l_{\widehat{v}}^{(k)}}}{H\Big(\Big(\overrightarrow{l_{u}^{(k)}}\Big)^{\pi_{v}}\Big)}\Bigg)^{\pi_{v}}\Bigg)} = \tau_{\widehat{v}}^{(k-1)} \label{eq:traversal-passive-backward}
		\end{split}
	\end{align}
\end{subequations}

\subsection{Genesis Block}
As the genesis block is the first block belonging to the user.
It is not meant to traverse backward from that block.
Hence, only the forward traversal information is sufficient, and we don't include the $\overleftarrow{l_{u}^{(0)}}$ and $\overleftarrow{l_{\widehat{v}}^{(0)}}$ as we have done in other blocks.
Also, as the genesis block does not describe any actual access event it does not follow the usual semantics of active and passive users. Rather the same user is simultaneously considered active and passive.
The Equations~\ref{eq:construction-genesis-active} and \ref{eq:construction-genesis-passive} describe the formulation of the active and the passive parts of the genesis block.

\resizebox{0.97\linewidth}{!}{%
\begin{minipage}{0.51\linewidth}
	\begin{subequations}
		\begin{align}
			\begin{split}
				\overrightarrow{l_{u}^{(0)}} =& g^{r_{t}^{(0)}}
			\end{split} \\
			\begin{split}
				l_{u}^{(0)*} =& H\Big(g^{w \pi_{t} r_{t}^{(0)}} g^{\pi_{t}}\Big)
			\end{split}
		\end{align}
		\label{eq:construction-genesis-active}
	\end{subequations}
\end{minipage}%
\begin{minipage}{0.52\linewidth}
	\begin{subequations}
		\begin{align}
			\begin{split}
				\overrightarrow{l_{\widehat{u}}^{(0)}} =& g^{r_{t}^{(0)}}
			\end{split} \\
			\begin{split}
				l_{uw}^{(0)} =&  H(g^{wr_{t}^{(0)}}) g^{h\pi_{t}r_{t}^{(0)}}
			\end{split}
		\end{align}
		\label{eq:construction-genesis-passive}
	\end{subequations}
\end{minipage}
}

Genesis blocks do not have addresses. 
The id of the block is computed as a hash of the public key of the user.

\subsection{Construction}
\label{sec:ledger-construction}
Blocks are constructed by the TS through active participation with the Active user $A_{u}$. 
The Request for Sensitive Information (RSI) contains information regarding the requested operation. 
The TS constructs the block registering the event and returns the information requested. 
While retrieving that information the TS gets to know about the patient who is the passive user $A_{v}$ in this context. 
With the knowledge of $A_{u}$, $A_{v}$, and some private information computed by the $A_{u}$ the TS constructs the block.

\begin{figure}
    \centering
    \includegraphics[width=0.3\textwidth]{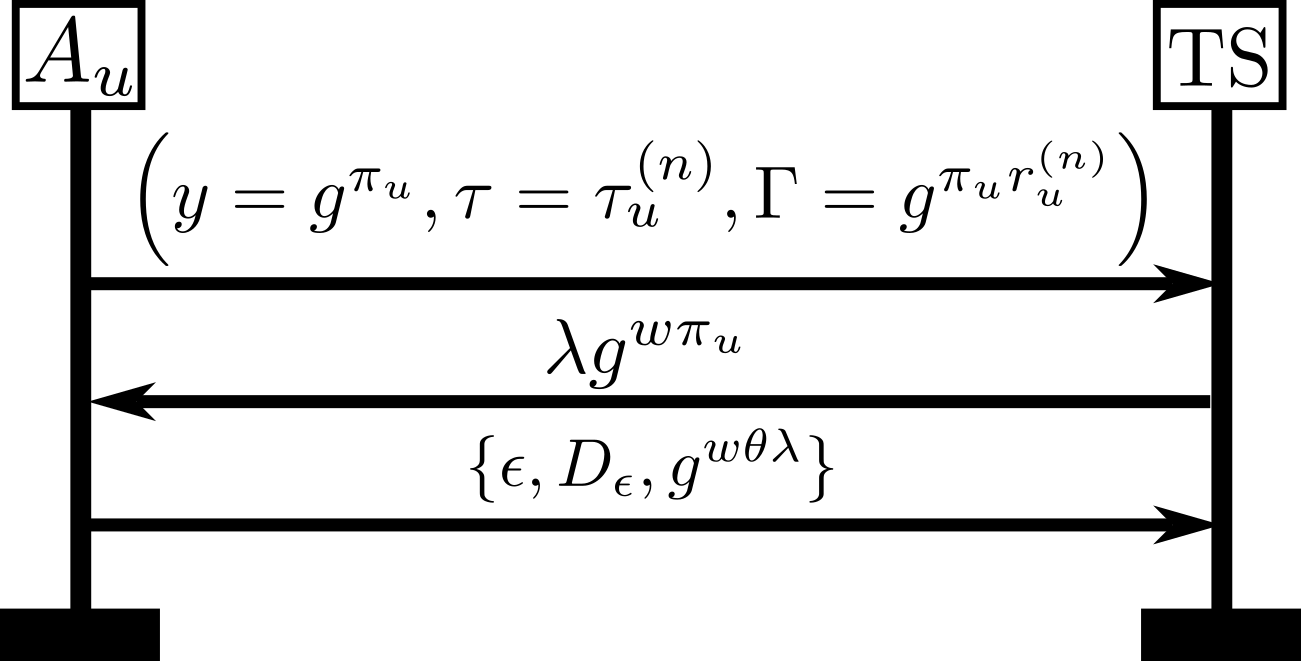}
    \caption{Protocol for Communicating RSI}
    \label{fig:protocol}
\end{figure}

We propose a two-stage protocol (shown in Figure~\ref{fig:protocol}) through which the TS obtains sufficient information securely without damaging the privacy of $A_{u}$.
In the first stage of the protocol, the active user claims to be $A_{u}$ by sending the public key $g^{\pi_{u}}$ and the last block $\tau_{u}^{(n)}$ in which $A_{u}$ was active.
It also sends a token $\Gamma$ calculated as $\Big(\overrightarrow{l_{u}^{(n)}}\Big)^{\pi_{u}}$ using which the TS verifies that claim of identity.
TS computes $H(y \Gamma^{w})$ and compares that with $l_{u}^{(n)*}$.
If they are the same, then the block $\tau_{u}^{(n)}$ is a block in which $A_{u}$ participated as an active user and the claim of identity is also correct because it has access to the private key of $A_{u}$.
To verify whether $\tau_{u}^{(n)}$ is the last such block TS computes $\tau_{u}^{(n)}\Gamma = c_{u}^{(n+1)}$ and checks whether it already exists in the index of addresses.
If the computed $c_{u}^{(n+1)}$ is not found and the claim of identity is proven, then the user is not malicious and the protocol follows the next stage. 
Otherwise, TS rejects the request.

In the second stage of the protocol, the active user communicates the intended operation and related data along with the access key, using which the TS recomputes $g^{\theta}$, which was lost initially.
The TS generates a random $\lambda \in Z_{q}^{*}$, computes its inverse $\lambda^{-1}$ and sends $\lambda\big(g^{\pi_{u}}\big)^{w} = \lambda g^{w\pi_{u}}$ to the $A_{u}$.
$A_{u}$ can extract the $\lambda$ by computing $\big(g^{w}\big)^{\pi_{u}}$ and then sends $g^{\theta \lambda w}$ calculated using inverse of its private key $(g^{\theta\pi_{u}w})^{\lambda\pi_{u}^{-1}}$.

The tuple in the third message shown in Figure~\ref{fig:protocol} summarizes the message that the Active user sends to the TS in stage 2.
The $\epsilon$ and $D_{\epsilon}$ denote the action that the active user intends to perform and the corresponding data respectively. The set of possible actions and their corresponding data are represented in Table~\ref{table:actions}.

\begin{table}[]
	\centering
	\begin{tabularx}{\linewidth}{|l|l|X|}
		\toprule
		$\epsilon$ & $D_{\epsilon}$                        & Intention                                                                                                     \\ \midrule
		identify & $a_{v,j}$                           & Retrieve public key of the patient associated with medical information~$a_{v_{j}}$                        \\ \midrule
		fetch    & $g^{\pi_{v}}$                       & Fetch all records of patient identified by public key \\ \midrule
		insert   & $g^\pi_{v},A$ & Insert records $A = \{a_{v,1}\dots a_{v,m}\}$ and associate them with patient identified by $g^{\pi_{v}}$\\ \midrule
		delete   & $a_{v,j}$                           & Delete Record $a_{v,j}$  \\
		\bottomrule
	\end{tabularx}
	\caption{Actions in Access Event}
	\label{table:actions}
\end{table}

After receiving the response, the TS uses the $\lambda^{-1}$ and $w^{-1}$ to reconstruct $g^{\theta}$. 
This $g^{\theta}$ is the secret using which the database anchors are encrypted.
TS can get the public key of the passive user $g^{\pi_{v}}$ either from $D_{w}$ or from the database using $g^{\theta}$. 
So, at this stage of the protocol, TS is aware of the passive user too.

Now the TS has enough information to construct the next block $\tau_{u}^{(n+1)}$ which is also $\tau_{v}^{(n+1)}$ for the passive user $A_{v}$.
TS generates two random numbers, $r_{u}^{(n+1)}$ and $r_{v}^{(n+1)}$, such that there exist not multiplicative inverse in $\mathbb{Z}_{p-1}$ and the inequality in Equation~\ref{eq:inequality-random-generation} is satisfied.
\begin{equation}
	H_{2}\Big(g^{\pi_{v}r_{u}^{(n+1)}}\Big) \not\equiv H_{2}\Big(g^{\pi_{u}r_{u}^{(n)}}\Big) \; (mod\; 2)
	\label{eq:inequality-random-generation}
\end{equation}

The construction of the parts of block $\tau_{u}^{(n+1)}$ are summarized in Equations~\ref{eq:protocol-construction-active}, \ref{eq:protocol-construction-passive} and \ref{eq:protocol-construction-address}.

\resizebox{0.99\linewidth}{!}{%
	\begin{minipage}{1.06\linewidth}
		\begin{subequations}
			\begin{align}
				\begin{split}
					\overrightarrow{l_{u}^{(n+1)}} &\coloneqq g^{r_{u}^{(n+1)}}
				\end{split} \\
				\begin{split}
					\overleftarrow{l_{u}^{(n+1)}} &\coloneqq H\Big(y_{u}^{r_{v}^{(n+1)}}\Big)\overrightarrow{l_{u}^{(n)}}= H\Big(g^{\pi_{u}r_{v}^{(n+1)}}\Big)g^{r_{u}^{(n)}}
				\end{split} \\
				\begin{split}
					l_{u}^{(n+1)*} &\coloneqq H\Big( y_{u}^{wr_{u}^{(n+1)}} y_{u} \Big) = H\Big( g^{\pi_{u}wr_{u}^{(n+1)}} g^{\pi_{u}} \Big)
				\end{split}
			\end{align}
			\label{eq:protocol-construction-active}
		\end{subequations}
		\begin{subequations}
			\begin{align}
				\begin{split}
					\overrightarrow{l_{\widehat{v}}^{(n+1)}} &\coloneqq  g^{r_{\widehat{v}}^{(n+1)}}
				\end{split} \\
				\begin{split}
					\overleftarrow{l_{\widehat{v}}^{(n+1)}} &\coloneqq H\Big( y_{\widehat{v}}^{r_{u}^{(n+1)}} \Big) \overrightarrow{l_{\widehat{v}}^{(n)}} = H\Big(g^{\pi_{v}r_{u}^{(n+1)}} \Big) g^{r_{v}^{(n)}}
				\end{split} \\
				\begin{split}
					l_{vw}^{(n+1)}  &\coloneqq H\Big(y_{w}^{r_{v}^{(n+1)}}\Big) y_{v}^{h r_{v}^{(n+1)}} = H\Big(g^{wr_{v}^{(n+1)}}\Big) g^{\pi_{v}h r_{v}^{(n+1)}}
				\end{split}
			\end{align}
			\label{eq:protocol-construction-passive}
		\end{subequations}
		\begin{subequations}
			\begin{align}
				\begin{split}
					c_{u}^{(n+1)} &\coloneqq \tau_{u}^{(n)} \Gamma
				\end{split} \\
				\begin{split}
					c_{\widehat{v}}^{(n+1)} &\coloneqq \tau_{\widehat{v}}^{(n)}\Bigg(\frac{l_{vw}^{(n)}}{H\Big(\Big(\overrightarrow{l_{\widehat{v}}^{(n)}}\Big)^{w}\Big)}\Bigg)^{h^{-1}}
					= \tau_{\widehat{v}}^{(n)}\Big(g^{\pi_{v}r_{v}^{(k)}}\Big)
				\end{split}
			\end{align}
			\label{eq:protocol-construction-address}
		\end{subequations}
	\end{minipage}
}

\subsection{Encrypting Block Contents}
\label{sec:block-contents}

The block contents should be encrypted in such a way that they can be decrypted by only the $A_{u}$, $A_{v}$ involved in that block, and all supervisors.
We want that decryption also to happen offline so that it does not require any network communication.
The TS formulates a straight line primarily with two coordinates (shown in Equation~\ref{eq:content-encryption-coordinates}) that only the users $A_{u}$ and $A_{v}$ can compute.
Although the TS can compute those coordinates while construction, it loses the information it needs to reconstruct that again.
Once the straight line is constructed, it finds two random coordinates that satisfy that linear equation.
One of those coordinates is published with that block as plain text along with the $x$ value of the other.
The cryptographic hash of the $y$ value of the other random coordinate is used as a password to symmetrically encrypt the message.

\begin{align}
	d^{(u)} = \begin{bmatrix}
		H_{2}\Big(g^{\pi_{u}r_{u}^{(n)}}\Big) \\
		c_{v}
	\end{bmatrix}, & \quad\quad%
	d^{(v)} = \begin{bmatrix}
		H_{2}\Big(g^{\pi_{v}r_{u}^{(n+1)}}\Big) \\
		c_{u}
	\end{bmatrix}
	\label{eq:content-encryption-coordinates}
\end{align}

We generate two random coordinates $\gamma, \delta$ on that line and publish $\gamma$ and $\delta_{x}$ with the block.
The contents are encrypted using $H_{2}(\delta_{y})$.
Both active and passive users can compute either $d^{(u)}$ or $d^{(v)}$ and then interpolate a straight line using $\gamma$ which is available as a plaintext.
Then, the users put $x = \delta_{x}$ on that equation and calculate $y = \delta_{y}$.
Once $\delta_{y}$ is retrieved, its hash $H_{2}(\delta_{y})$ can be computed, using which the encrypted message can be decrypted and the plaintext can be obtained.

However, the supervisors cannot compute either $d^{(u)}$ or $d^{(v)}$, hence cannot use $\lambda$ to interpolate the straight line.
As the supervisors can also perform access events they too have their access key like the Data Managers.
Additionally, they have $g^{\phi w \pi_{s}}$ which is specifically used for viewing, but unlike $g^{\theta}$ it is not lost by the TS.
Although  $g^{\theta}$ is lost, it is reconstructed in the second stage of the protocol by the TS.
Hence, the TS computes a suffix $\Big(g^{\theta w} g^{\phi w}\Big)^{\gamma_{x}}$ and multiplies the $H_{2}(\delta_{y})$ as shown in Equation~\ref{eq:content-encryption-supervisor-secret} and stores that $\delta_{y}^{(s)}$ in the block.

\begin{equation}
	\delta_{y}^{(s)} = H_{2}(\delta_{y})\Big(g^{\theta w} g^{\phi w}\Big)^{\gamma_{x}} = H_{2}(\delta_{y}) g^{(\theta + \phi)w\gamma_{x}}
	\label{eq:content-encryption-supervisor-secret}
\end{equation}

The supervisors $A_{s}$ can compute the suffix and obtain $H_{2}(\delta_{y})$ by division in $\mathcal{Z}_{p}$ as shown in Equation~\ref{eq:content-encryption-supervisor-secret-retrival}.

\begin{equation}
	\frac{\delta_{y}^{(s)}}{ \Big( (g^{\phi w \pi_{s}})^{\pi_{s}^{-1}} (g^{\theta w \pi_{s}})^{\pi_{s}^{-1}}\Big)^{\gamma_{x}} }
	\label{eq:content-encryption-supervisor-secret-retrival}
\end{equation}

However, the $A_{u}$ and $A_{v}$ also can compute $H_{2}(\delta_{y})$ and can extract the suffix $g^{(\theta + \phi)w\gamma_{x}}$ through division. 
As $\gamma_{x}$ is known anyone can compute $\gamma_{x}^{-1}$ and and extract $g^{(\theta + \phi)w} = (g^{(\theta + \phi)w\gamma_{x}})^{\gamma_{x}^{-1}}$ using that.
Once extracted it can be reused with some other $\gamma_{x}$ associated with some other block.
Hence, while generating random $\gamma$ we need to ensure that $\nexists \gamma_{x}^{-1}$ such that $\gamma_{x} \gamma_{x}^{-1} \equiv 1 \in \mathcal{Z}_{(p-1)}$.

Finally a checksum of the block is calculated and the blocks are signed by the Trusted Server.

\section{Evaluation}
\label{sec:evaluation}

We perform three experiments to measure the performance of our proposed solution in terms of CPU time consumption.
 
\paragraph{Experimental Setup:} To test the performance of our proposed scheme we have implemented our cryptographic functions in C++ using Crypto++ library.
The Trusted Server has been implemented as a TCP Server using the `Boost Asio` library.
The program that takes the private key of the user and performs operations described in Table~\ref{table:actions} is implemented as a TCP client.
PostgreSQL database has been used for storing HD.
Redis key-value store has been used as an event log and for the index.
To measure the performance of active or passive traversal we implement a reader application that takes the private key of a user and performs traversal by accessing the key-value store.

We initialize the system with 5 managers, 4 supervisors, 7 patients and their genesis records in the database. 
In this paper, we do not focus on secure distribution of the private keys to the users. 
Hence, we assume that the private keys are securely delivered to them.

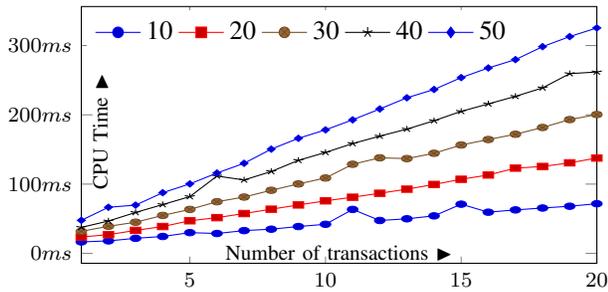
\begin{figure}[]
	\centering
	\begin{tikzpicture}
		\begin{axis}[
			legend columns=5,
			anchor=north west,
			xmin=1, xmax=20,
			xlabel={Number of transactions $\blacktriangleright$},
			ylabel={CPU Time $\blacktriangleright$},
			yticklabel={$\pgfmathprintnumber{\tick}ms$},
			y label style={at={(axis description cs:0.25,.8)},anchor=south},
			x label style={at={(axis description cs:.5,0.16)},anchor=south},
			x label style={font=\footnotesize},
			y label style={font=\footnotesize},
			ticklabel style={font=\footnotesize},
			yscale=0.6,
			legend style={draw=none},
			legend style={at={(0.01,1.3)},anchor=south west}
			]
			\addplot table [x=N, y=Time, col sep=comma] {results/10.csv};
			\addlegendentry{10}
			\addplot table [x=N, y=Time, col sep=comma] {results/20.csv};
			\addlegendentry{20}
			\addplot table [x=N, y=Time, col sep=comma] {results/30.csv};
			\addlegendentry{30}
			\addplot table [x=N, y=Time, col sep=comma] {results/40.csv};
			\addlegendentry{40}
			\addplot table [x=N, y=Time, col sep=comma] {results/50.csv};
			\addlegendentry{50}
		\end{axis}
	\end{tikzpicture}
	\caption{Bulk Insertion}
	\label{fig:results-insertion}
\end{figure}

\paragraph{Experiment 1}
First, we measure the CPU time consumed by the TS during the insertion operation by sending a series of bulk insertion requests for 5 patients. 
In each insertion operation, a batch of records is sent. 
We vary the batch size from 10 to 50 records per request in each iteration. 
So, at the end of $20^{th}$ request, there are 200 records associated with patient 1 and 50*20 =1000 associated with the $5^{th}$ patient. 
The results are shown in Figure~\ref{fig:results-insertion}. 
We observe that the time consumed for inserting records is not related to the total number of HD in the database. However, it is directly proportional to the number of existing records already associated with a patient.
Such a result is expected because the TS performs a linear traversal over the database while inserting a record. 
It is observed that the time consumption is more than 300ms when there are 1000 records associated with a single patient.

\begin{figure}[]
	\centering
	\begin{tikzpicture}
		\begin{axis}[
			legend columns=5,
			xmin=200, xmax=1001,
			xlabel={Number of Records $\blacktriangleright$},
			ylabel={CPU Time $\blacktriangleright$},
			yticklabel={$\pgfmathprintnumber{\tick}ms$},
			y label style={at={(axis description cs:0.27,1.3)},anchor=south},
			x label style={at={(axis description cs:.5,0.6)},anchor=south},
			x label style={font=\footnotesize},
			y label style={font=\footnotesize},
			ticklabel style={font=\footnotesize},
			yscale=0.32,
			]
			\addplot table [x=N, y=Time, col sep=comma] {results/fetch.csv};
		\end{axis}
	\end{tikzpicture}
	\caption{Fetch All Entries}
	\label{fig:results-fetch-all}
\end{figure}
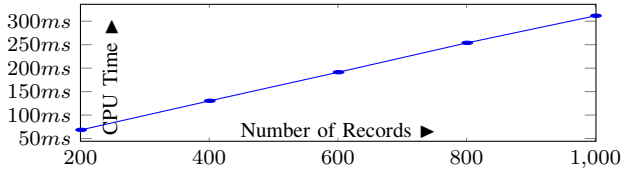

\begin{figure}[]
	\centering
	\begin{tikzpicture}
		\begin{axis}[
			legend columns=2,
			xmin=50, xmax=251,
			xlabel={Number of Entries $\blacktriangleright$},
			ylabel={CPU Time $\blacktriangleright$},
			yticklabel={$\pgfmathprintnumber{\tick}ms$},
			y label style={at={(axis description cs:0.25,.7)},anchor=south},
			x label style={at={(axis description cs:.5,0.2)},anchor=south},
			x label style={font=\footnotesize},
			y label style={font=\footnotesize},
			ticklabel style={font=\footnotesize},
			yscale=0.56,
			legend style={draw=none, fill=none},
			legend style={at={(0.01,1.1)},anchor=south west}
			]
			\addplot table [x=N, y=Time, col sep=comma] {results/traversal/active-forward.csv};
			\addlegendentry{Active Forward}
			\addplot table [x=N, y=Time, col sep=comma] {results/traversal/active-backward.csv};
			\addlegendentry{Active Backward}
			\addplot table [x=N, y=Time, col sep=comma] {results/traversal/passive-forward.csv};
			\addlegendentry{Passive Forward}
			\addplot table [x=N, y=Time, col sep=comma] {results/traversal/passive-backward.csv};
			\addlegendentry{Passive Backward}
		\end{axis}
	\end{tikzpicture}
	\caption{Traversal}
	\label{fig:results-traversal}
\end{figure}
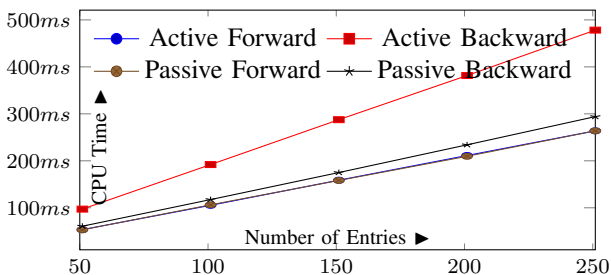

\paragraph{Experiment 2}
Next, we fetch all records associated with a single patient and measure the CPU time consumed. 
The results shown in Figure~\ref{fig:results-fetch-all} suggest that the time consumed for retrieving HD is directly proportional to the number of records associated with a patient.

\paragraph{Experiment 3}
In this experiment, we check the performance of ledger traversal.

We perform several traversal requests to the ledger each limiting the number of entries from 50 to 250.
The results are shown in Figure~\ref{fig:results-traversal}.
We observe both active and passive forward traversal have the same performance characteristics.
Although the Passive traversal is slightly more expensive than the Forward ones, the Active Backward traversal is more expensive than the Passive one.

\section{Conclusion and Discussion}
\label{sec:conclusion}
In this paper, we have proposed a secure, responsible, privacy-preserving document storage and retrieval technique.
The solution can be used for different business processes other than healthcare.
Even in healthcare, the applications of such systems are not limited to health registries.
However, a generalized use case may involve more than two users in one event.

Furthermore, although the cost of the insertion operation does not depend on the number of total records in the database, it depends on the number of records associated with one patient. 
Therefore, it is suitable for scenarios where the per-patient record is lower than the total number of records.
Also, the traversability of the blocks in the ledger cannot be verified by any entity because that entity can also partition the ledger into blocks associated with one particular user or the other.
Not only that, but also the verifier will be able to understand the sequence of events associated with any user inside the system.
This leads to another important problem that future work may address.
In our proposed solution, the immutability of the ledger is maintained by annotating an entry with the hash of the previous entry. 
However, it is a centralized system implemented using a key-value store.
An implementation using blockchain-based distributed ledger platforms may provide byzantine fault tolerance against manipulation of the ledger.

\section*{Acknowledgment}
This work is supported by the Research Council of Norway, grant number 288106. 
We also acknowledge Jan F. Nygård from the Cancer Registry of Norway (CRN) for discussing the internal operations of CRN regarding this problem.

\bibliographystyle{IEEEtran}
\bibliography{main}

\end{document}